\documentclass{appolb}
\usepackage{epsfig}
\newcommand {\pom} {I\!\!P}
\newcommand {\pomsub} {{\scriptscriptstyle \pom}}
\newcommand {\xpom} {x_{\pomsub}}
\newcommand{\dkap}{\Delta\kappa^{\gamma}}

\newcommand{\lam}{\lambda^{\gamma}}
\newcommand{\wwgamma}{WW\gamma}

\begin{document}
\title{Diffraction at HERA, Tevatron and LHC
\thanks{Presented at the 2009 Epiphany Conference, Cracow, in honour of Jan
Kwiecinski}%
}
\author{Christophe Royon
\address{IRFU-SPP, CEA Saclay, F91 191 Gif-sur-Yvette, France}
}
\maketitle
\begin{abstract}
In this short review, we describe some of the results on diffraction from 
the Tevatron and give some prospects for the LHC. In particular, 
we discuss the search for exclusive events at the Tevatron and their
importance for the LHC program/ We finish by presenting
the project of installing forward detectors in the ATLAS collaboration at 220
and 420 m.
\end{abstract}
  
\section{Inclusive diffraction at HERA}
In this short review, we will only describe briefly the measurement of the inclusive
diffractive structure function $F_2^{D}$ and the extraction of the gluon density
of the Pomeron. A more detailled study can be found in Ref.~\cite{ourreview}.

\begin{figure}
\begin{center}
\vspace{10.cm}
\hspace{-5cm}
\epsfig{file=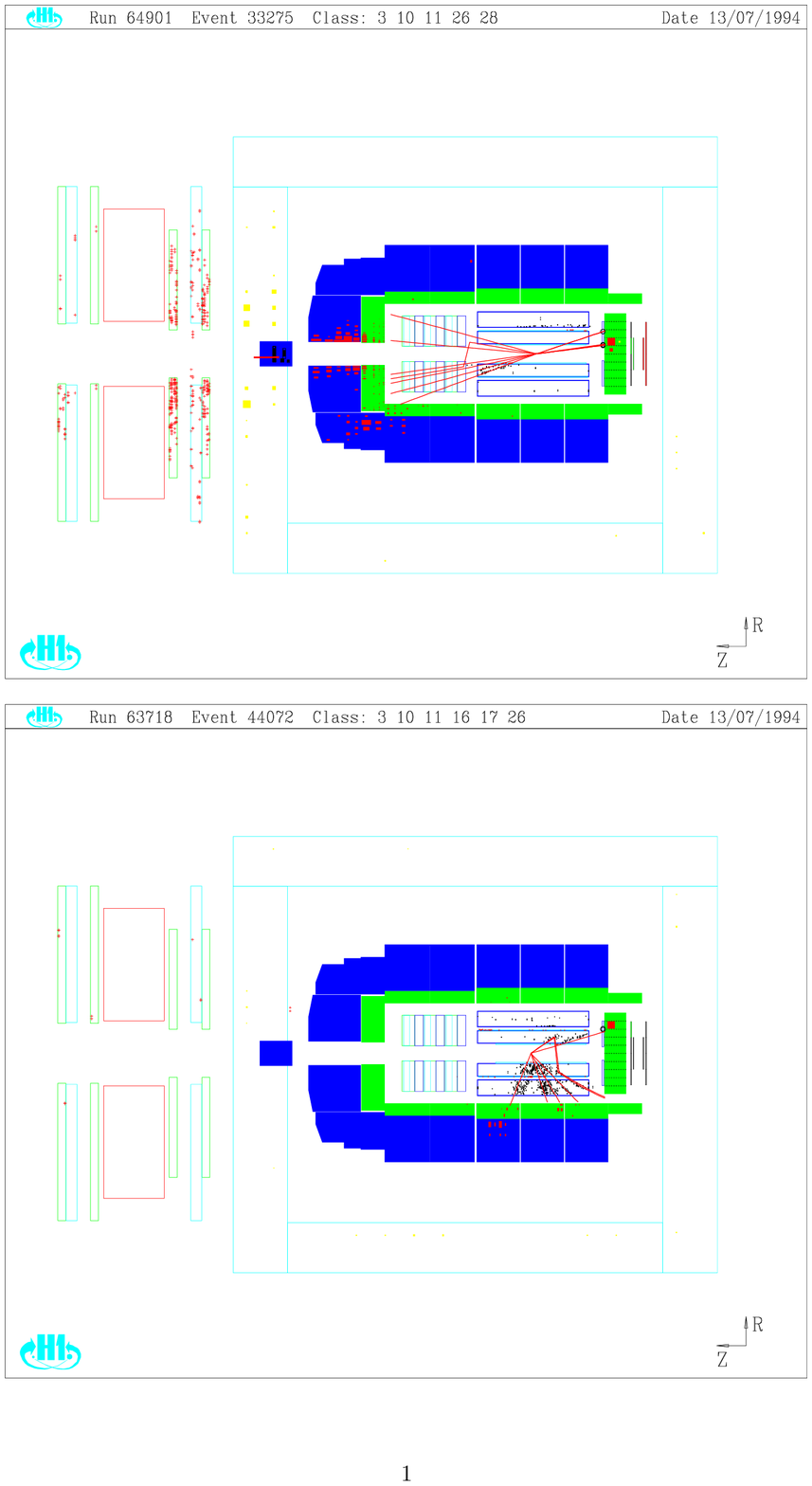,width=5.2cm}
\caption{``Usual" and diffractive events in the H1 experiment.}
\label{fig1}
\end{center}
\end{figure}

In Fig.~\ref{fig1}, we give an event display of a
standard deep inelastic interaction (DIS) event (top) and from a diffractive one
(botttom). For typical DIS events, the electron is scattered in the backward 
detector or
the LAr calorimeter and the proton is destroyed. Part of the
proton energy can be detected in the forward detectors such as the forward
liquid argon calorimeter, the PLUG calorimeter or the forward muon detector.
In about 10\% of the events (see Fig.~\ref{fig1}, bottom), the situation is
different: there is no energy in the forward part of the detector. It means that
there is no colour exchange between the proton and the jet produced in the
event. In most of these events, the proton remains intact and is scattered at
very small angle from the beam direction.
This brings us to two different possibilities of detecting diffractive events,
namely the rapidity gap selection where a gap devoid of energy is requested in
the forward region of the detector, and the proton tagging detection where
special detectors can be placed close to the beam in the very forward direction
\footnote{Another method developped by the ZEUS collaboration is based on the fact
that non diffractive events are exponentially suppressed at high values of $M_X$,
the total invariant mass produced in the event.}.

\begin{figure}
\begin{center}
\epsfig{file=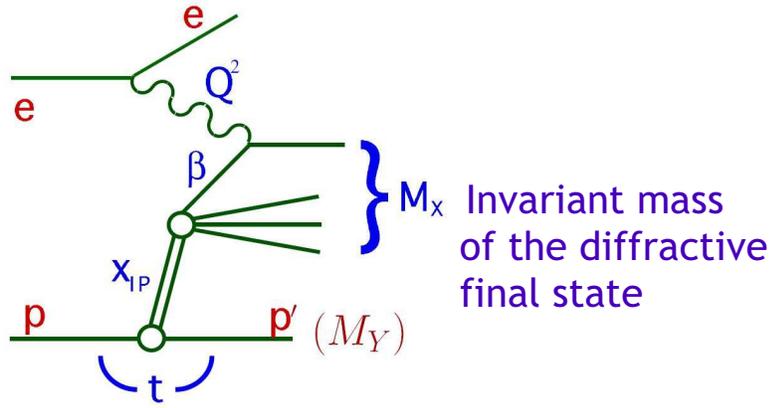,width=6cm,angle=270}
\caption{Scheme of a diffractive event at HERA.}
\label{fig2}
\end{center}
\end{figure}

The scheme of a diffractive event is shown in Fig.~\ref{fig2}. 
In order to describe the diffractive processes
where there is no colour exchange between the proton in the final state and the
scattered jet,
we have to introduce new variables in addition to the ones used to
describe the inclusive DIS such as $Q^2$, $W$, $x$ and $y$. Namely,
we define $\xpom$, which is the
momentum fraction of the proton carried by the colourless object (called the
pomeron), and $\beta$, the momentum fraction of the pomeron carried by the
interacting parton inside the pomeron, if we assume the pomeron to be made of
quarks and gluons
\begin{eqnarray}
\xpom &=& \xi = \frac{Q^2+M_X^2}{Q^2+W^2} \\
\beta &=& \frac{Q^2}{Q^2+M_X^2} = \frac{x}{\xpom}.
\end{eqnarray}

In the same way as the proton structure function is measured at HERA using DIS
interactions, it is possible to measure the diffractive structure function for
diffractive events~\cite{heraf2d}. Using the same analogy, it was proposed to
perform a QCD DGLAP~\cite{dglap} (Dokshitzer Gribov Lipatov Altarelli Parisi)
fit to the $F_2^D$ data to extract the quark and gluon densities in the Pomeron
if collinear and Regge factorisation are assumed~\cite{heraf2d,f2dfit}. According to 
Regge theory, we can factorise the 
$(\xpom,t)$ dependence from the $(\beta,Q^2)$ one for each trajectory
(Pomeron and Reggeon). The diffractive structure
function then reads:
\begin{eqnarray}
F_2^D \sim f_p(\xpom) (F_2^D)_{Pom}(\beta, Q^2) + 
f_r(\xpom) (F_2^D)_{Reg}(\beta, Q^2)
\end{eqnarray}
where $f_p$ and $f_r$ are the pomeron and reggeon fluxes, and $(F_2^D)_{Pom}$
and $(F_2^D)_{Reg}$ the pomeron and reggeon structure functions. The flux
parametrisation is predicted by Regge theory.
The DGLAP QCD fit allows to obtain the parton distributions in the Pomeron as a
direct output of the fit, and the gluon density is found to be much
higher than the quark one, showing that the Pomeron is gluon dominated. 
The gluon density at high $\beta$ is poorly constrained~\cite{heraf2d,f2dfit}.

\section{Diffraction at Tevatron}
At the Tevatron,
diffraction can occur not only on either $p$ or $\bar{p}$ side as at 
HERA, but also on both sides. In the same way
as the kinematical variables $\xpom$ and $\beta$ are defined
at HERA, we define $\xi_{1,2}$(=$\xpom$ at HERA) 
as the proton fractional momentum loss (or as the $p$ or
$\bar{p}$ momentum fraction carried by the pomeron), and $\beta_{1,2}$, the fraction of the
pomeron momentum carried by the interacting parton. The produced diffractive
mass is equal to $M^2= s \xi_1 $ for single diffractive events and to
$M^2= s \xi_1 \xi_2$ for double pomeron exchange where
$\sqrt s$ is the center-of-mass energy. The size of the rapidity gap
is of the order of $\Delta \eta \sim \log 1/ \xi_{1,2}$.

\subsection{Factorisation breaking}
A natural question to ask is whether one can use the diffractive PDFs
extracted at HERA to describe hard diffractive processes in hadron-hadron
collisions, and especially to predict the
production of jets, heavy quarks or weak gauge bosons at the Tevatron.  

\begin{figure}
\begin{center}
\epsfig{file=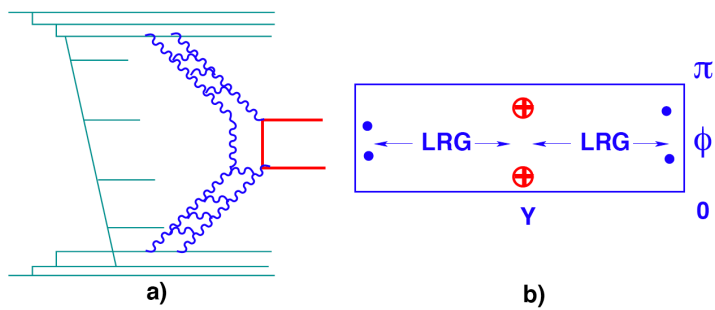,width=10cm}
\caption{Concept of survival probability.}
\label{fig3}
\end{center}
\end{figure}

From a theoretical point of view, diffractive hard-scattering 
factorization does not apply to
hadron-hadron collisions because of additional interactions between the
particles in initial and final states, as illustrated in Fig.~\ref{fig3}.
It is also worth noticing that the time scale when
factorisation breaking occurs is completely different from the hard interaction
one. Factorisation breaking is due to soft exchanges occuring in the initial and
final states which appear at a much longer time scale than the hard 
interaction. In that sense, it is expected that the survival probability,
defined as the
probability that there is no soft additional interaction or in other words that
the event remains diffractive, will
not depend strongly on the type of hard interaction and its kinematics. In other words,
the survival probability should be similar if one produces jets of different
energies, vector mesons, photons, etc, which can be cross checked experimentally
at Tevatron and LHC.

\subsection{Diffractive exclusive events and their interest at LHC}
\begin{figure}
\begin{center}
\epsfig{file=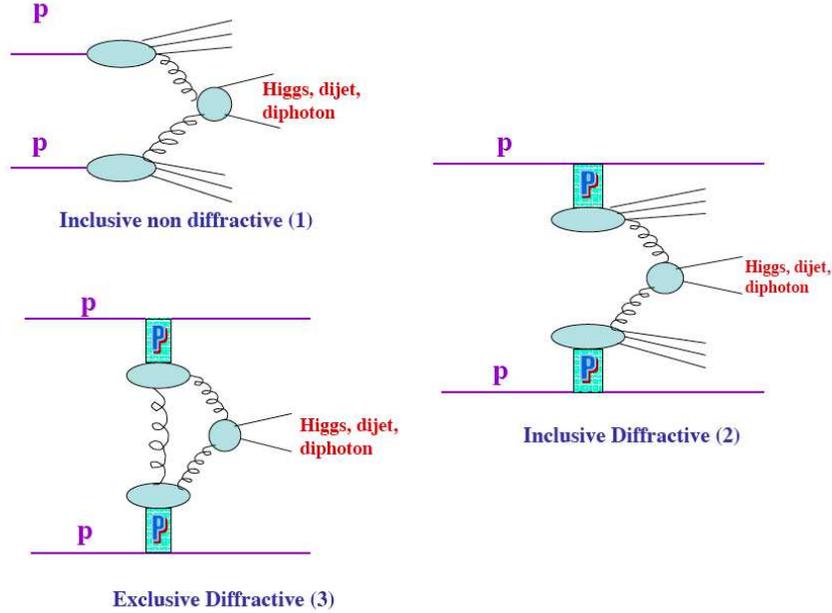,width=12cm}
\caption{Scheme of non diffractive, inclusive double pomeron exchange,
exclusive diffractive events at the Tevatron or the LHC.}
\label{fig7}
\end{center}
\end{figure}

A schematic view of non diffractive, inclusive double pomeron exchange,
exclusive diffractive events at the Tevatron or the LHC is displayed in
Fig.~\ref{fig7}.
The upper left plot (1) shows the ``standard" non diffractive events
where the Higgs boson, the dijet or diphotons are produced directly by a
coupling to the proton and shows proton remnants. The right plot (2) displays 
the standard diffractive double
pomeron exchange where the protons remain intact after interaction and the total
available energy is used to produce the heavy object (Higgs boson, dijets,
diphotons...) and the pomeron remnants. There is a third class of processes
displayed in the lower left figure (3), namely the exclusive diffractive
production. In this kind of events, the full energy is used to produce the heavy
object (Higgs boson, dijets, diphotons...) and no energy is lost in pomeron
remnants. 

There is an important consequence for the diffractive exclusive events: 
the mass of the
produced object can be computed using forward detectors and tagged
protons~\footnote{The formula is more complicated for low mass objects when the
proton mass cannot be neglected~\cite{chic}.}
\begin{eqnarray}
M = \sqrt{\xi_1 \xi_2 s}
\end{eqnarray} 
where $\sqrt{s}$ is the center-of-mass energy and $\xi$ is the fraction of the
proton momentum carried away by the Pomeron (called $\xpom$ at HERA).
The advantage of those processes is obvious: we can benefit from the
good forward detector resolution on $\xi$ to get a good mass resolution, and to
measure precisely 
the kinematical properties of the produced
object.

\subsection{Search for diffractive exclusive events at the Tevatron}
\begin{figure}
\begin{tabular}{cc}
\hspace{-1cm}
\epsfig{figure=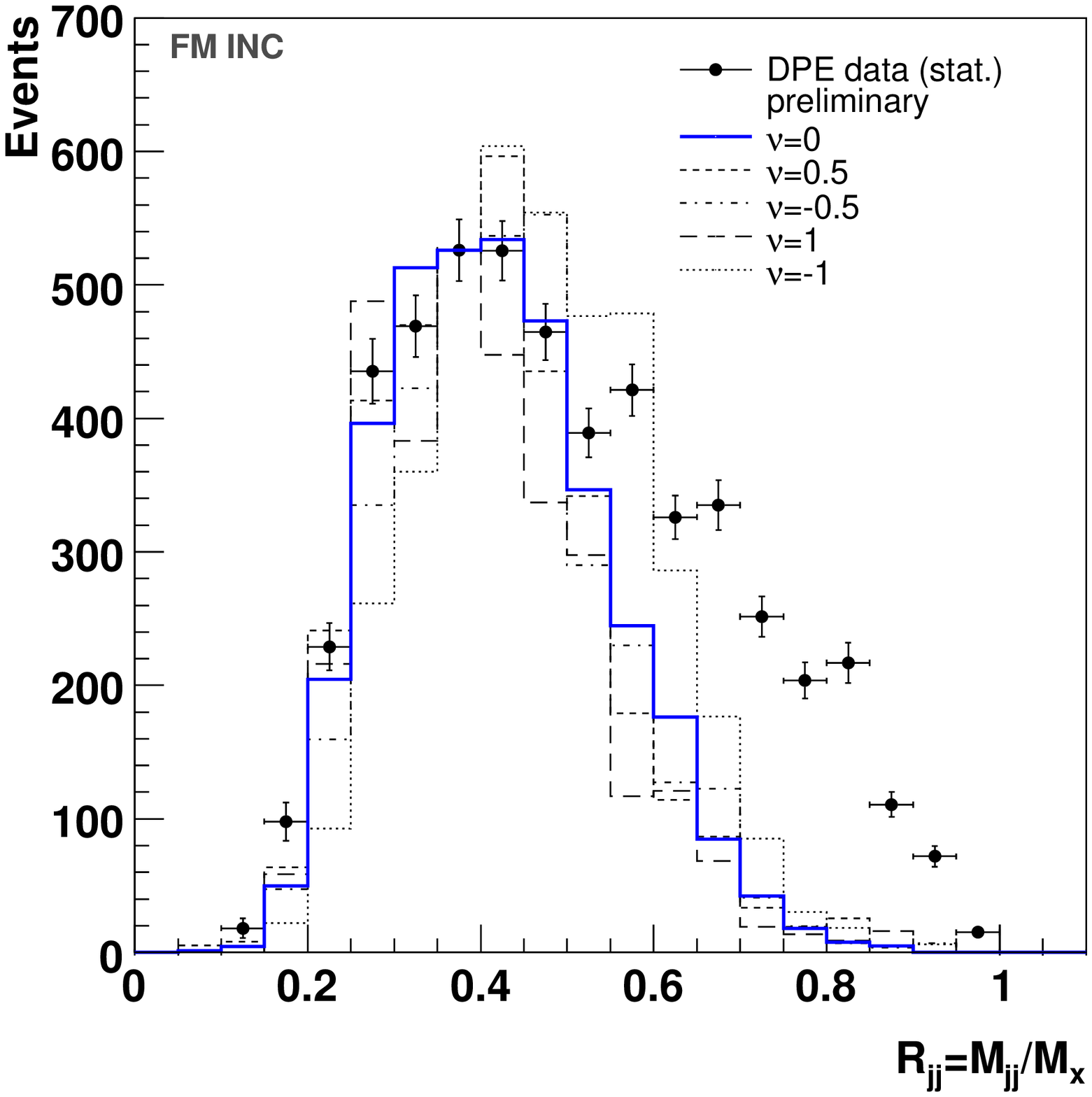,height=2.5in}  &
\epsfig{figure=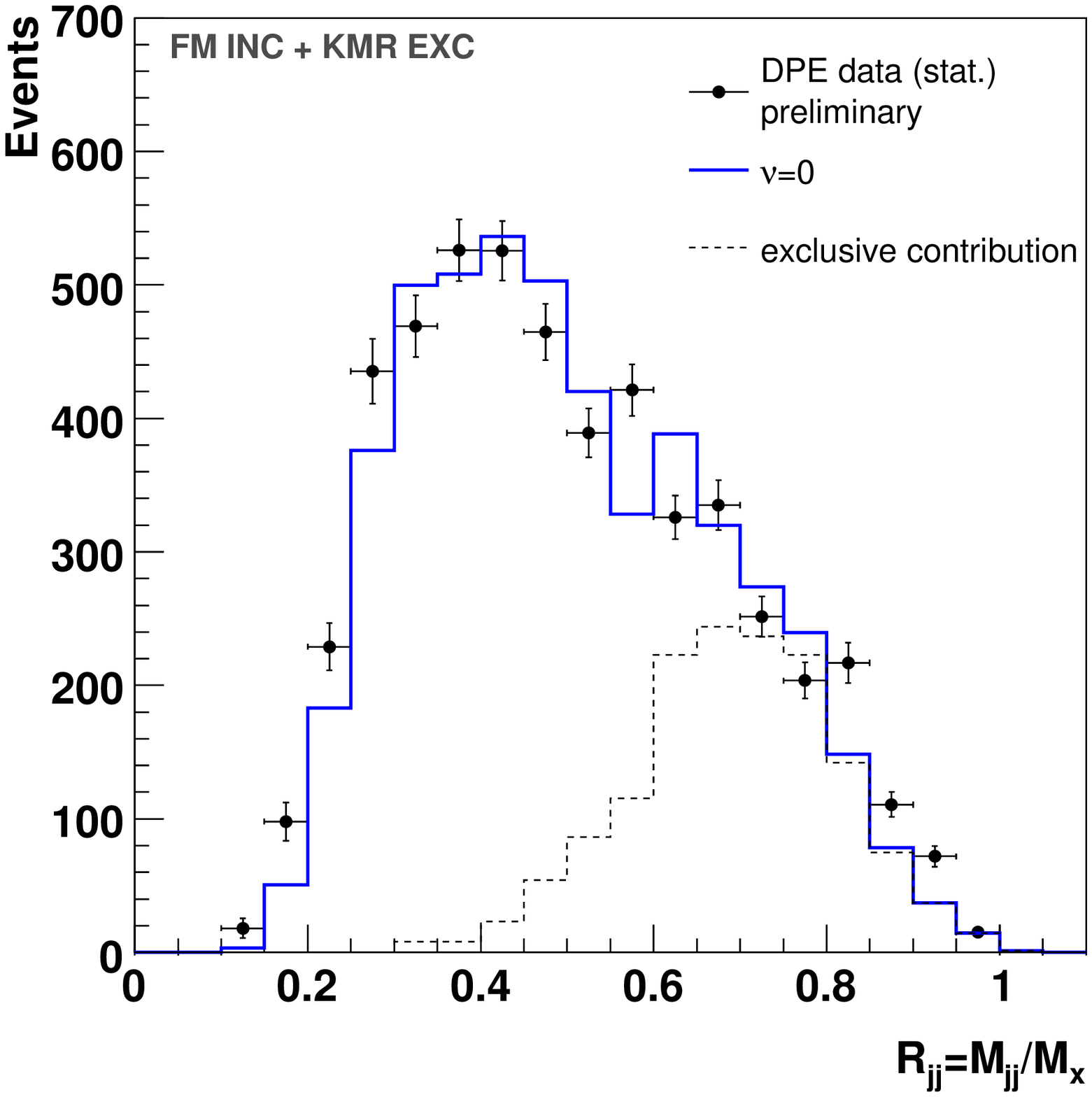,height=2.5in} \\
\end{tabular}
\caption{Left: Dijet mass fraction measured by the CDF collaboration compared to the
prediction from inclusive diffraction based on the parton densities
in the pomeron measured at HERA. The gluon
density in the pomeron at high $\beta$ was modified by varying the parameter
$\nu$.
Right: Dijet mass fraction measured by the CDF collaboration compared to the
prediction adding the contributions from inclusive and exclusive diffraction.}
\label{compare2}
\end{figure}

The CDF collaboration measured the so-called dijet mass fraction
in dijet events --- the ratio of the mass carried by the two jets produced in the event 
divided by the
total diffractive mass --- when the antiproton is tagged in the roman pot
detectors and when there is a rapidity gap on the proton side to ensure that the
event corresponds to a double pomeron exchange~\cite{cdfrjj}. 
This measurement is compared 
to the expectation obtained from the pomeron structure in quarks and gluons as
measured at HERA~\cite{f2dfit} 
(the factorisation breaking between
HERA and the Tevatron is assumed to be constant and to
come only through the gap survival probability,
0.1 at the Tevatron). 
The comparison between the CDF data for a jet $p_T$ cut of 10 GeV as an
example and the predictions from inclusive diffraction is given in 
Fig.~\ref{compare2}, left, together with
the effects of changing the gluon density at high $\beta$ by
changing the value of the $\nu$ parameter. Namely, to study the 
uncertainty on the gluon density at high $\beta$, 
we multiply the gluon distribution by the
factor $(1 - \beta)^{\nu}$. The $\nu$ parameter
varies between -1 and 1 (for $\nu=-$1 (resp. $+$1), the gluon density in the pomeron is
enhanced (resp. damped) at high $\beta$). 
QCD fits to the H1 data lead to 
an uncertainty on the $\nu$ parameter of 0.5~\cite{f2dfit}. Inclusive
diffraction alone is not able to describe the CDF data at high dijet mass fraction,
where exclusive events are expected to appear~\cite{oldab}. The conclusion
remains unchanged when jets with $p_T>25$ GeV are considered.

Adding exclusive events to the distribution of the dijet mass fraction leads to
a good description of data~\cite{oldab} as shown in Fig.~\ref{compare2}, right. This 
does not prove that exclusive events exist but shows that some 
additional component with respect to
inclusive diffraction compatible with exclusive events is needed to explain CDF data. 
To be sure of the existence
of exclusive events, the observation will have to be done in different channels
and the different cross sections to be compared with theoretical expectations.

Another interesting observable in the dijet channel is to look at the rate
of $b$ jets as a function of the dijet mass fraction. In exclusive events, the
$b$ jets are suppressed because of the $J_Z=0$ selection rule~\cite{ushiggs},
and as
expected, the fraction of $b$ jets in the diffractive dijet sample
diminishes as a function of the dijet mass fraction~\cite{cdfrjj}). 

The CDF collaboration also looked for the exclusive production of dilepton and
diphoton~\cite{cdfgamma}. 
Contrary to diphotons, dileptons cannot be produced exclusively via pomeron exchanges since
$g g \rightarrow \gamma \gamma$ is possible, but $g g \rightarrow l^+ l^-$ 
directly is impossible. Dileptons are produced via QED processes, and
the CDF dilepton measurement is $\sigma = 1.6
^{+0.5}_{-0.3} (stat) \pm 0.3 (syst)$ pb which is found to be in good agreement
with QED predictions. 3 exclusive diphoton events have been observed by the CDF
collaboration leading to a cross section of
$\sigma = 0.14
^{+0.14}_{-0.04} (stat) \pm 0.03 (syst)$ pb compatible with the expectations
for exclusive diphoton production at the Tevatron. Unfortunately, the number of events
is very small and the conclusion concerning the
existence of exclusive events is uncertain. An update by the CDF collaboration with higher luminosity
is however expected very soon. This channel will be however very
important at the LHC where the expected exclusive cross section is much higher.

\section{Exclusive diffraction at LHC}

\subsection{Exclusive Higgs production at the LHC}

One special interest of
diffractive events at the LHC is related to the existence of exclusive events
and the search for Higgs bosons at low mass in the diffractive mode.
So far, two projects are being discussed at the LHC: the installation of 
forward detectors at 220 and 420 m in ATLAS and CMS~\cite{afp} which we 
describe briefly at the end of this review.

\begin{figure}
\begin{center}
\begin{tabular}{cc}
\hspace{-1cm}
\epsfig{figure=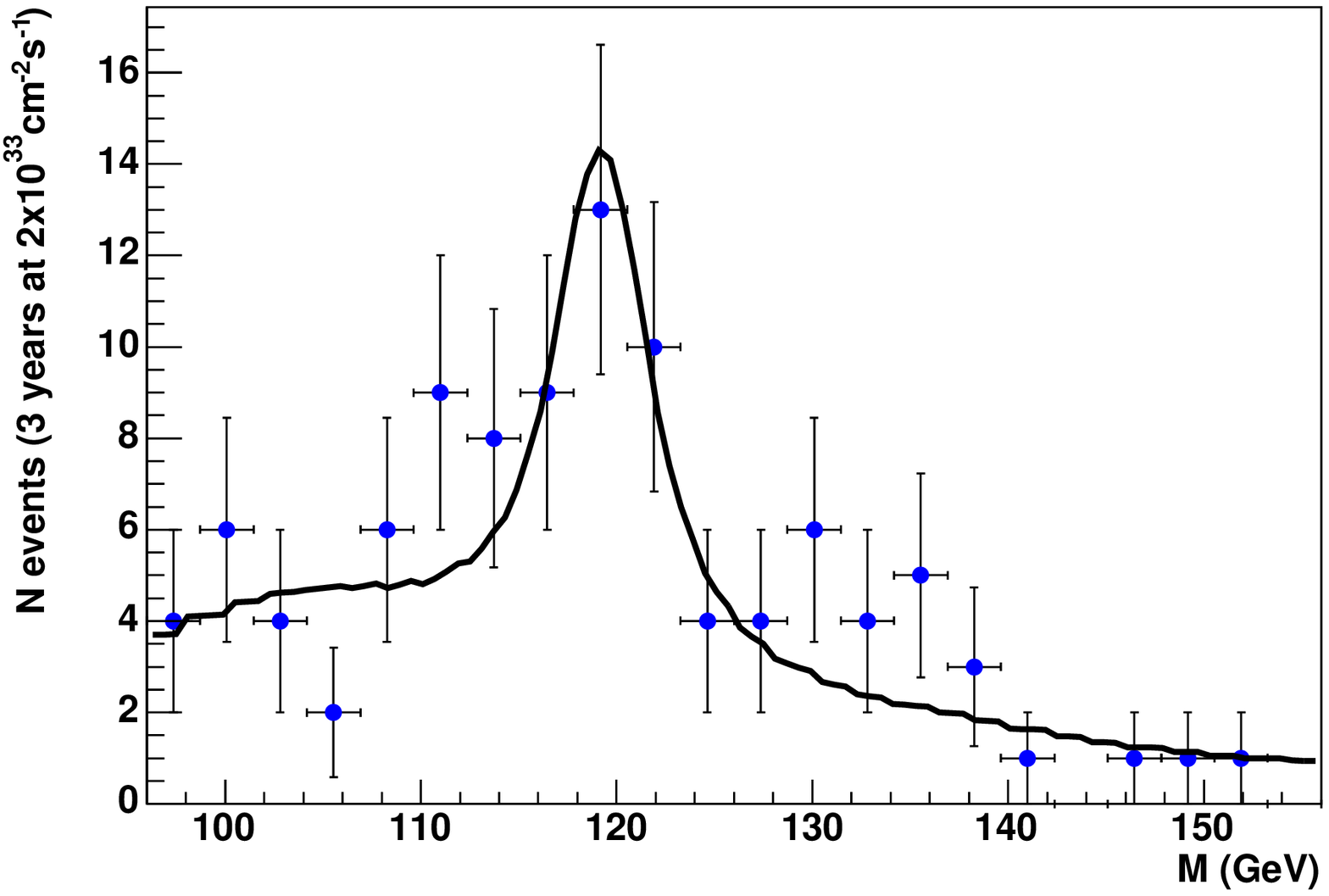,height=1.8in} &
\epsfig{figure=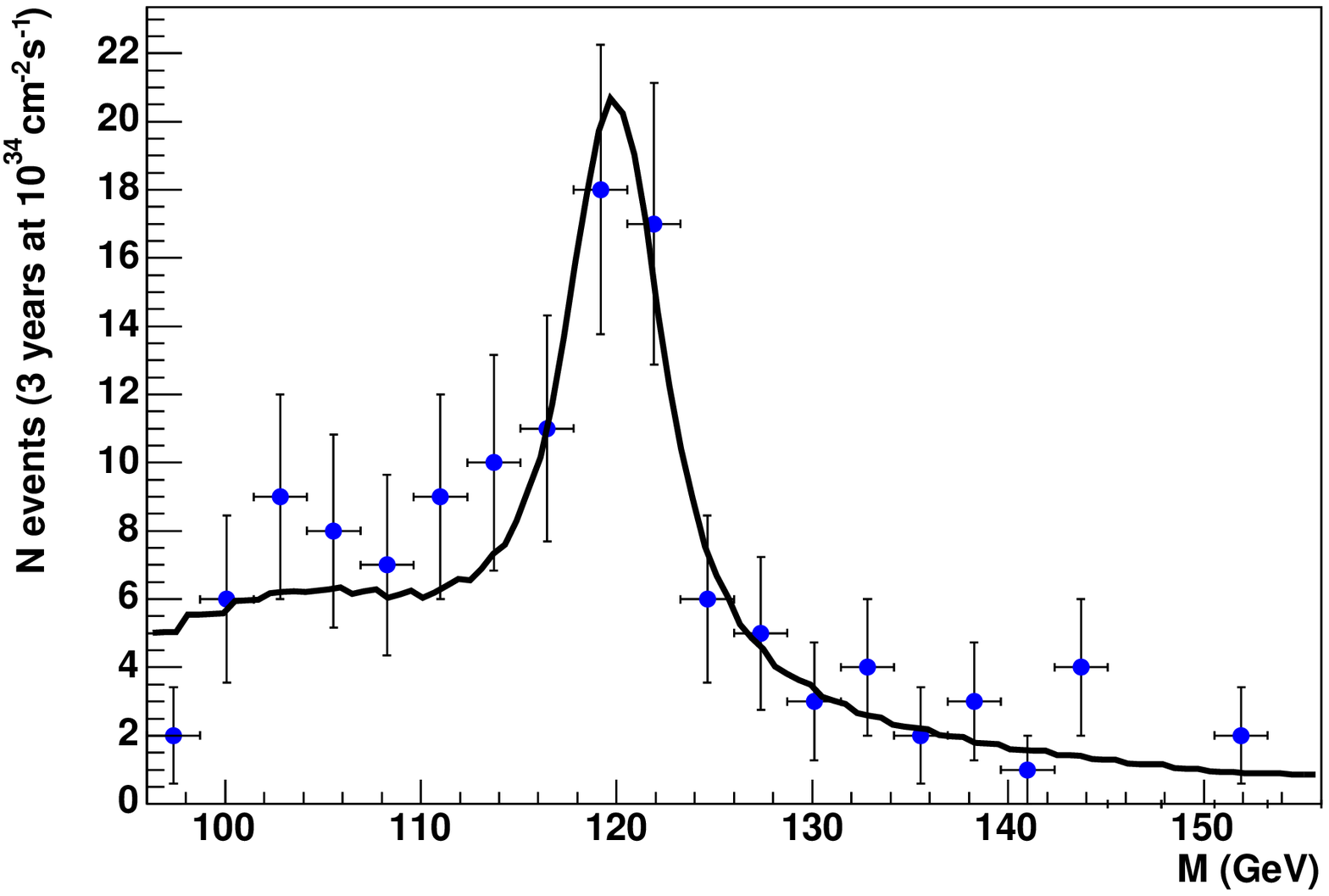,height=1.8in} \\
\end{tabular}
\caption{Higgs signal and background obtained for MSSM Higgs production
for neutral light CP-even Higgs bosons. 
The signal significance is larger than 3.5 $\sigma$ for 60 fb$^{-1}$
(left plot) and larger than 5 $\sigma$ after three years of data taking at high
luminosity at the LHC and using timing detectors with a resolution of 2 ps
(right plot).}
\label{signif}
\end{center}
\end{figure}

Many studies (including pile up effects and all background sources
for the most recent ones) 
were performed
recently~\cite{ushiggs,lavignac,higgsnew} to study in detail the signal 
over background for
MSSM Higgs production in particular. 
In Fig.~\ref{signif}, we give the number of  background and MSSM Higgs signal events
for a Higgs mass of 120 GeV for $\tan \beta \sim$40.
The signal significance is larger than 3.5 $\sigma$ for 60 fb$^{-1}$
(see Fig.~\ref{signif} left) and larger than 5 $\sigma$ after three years of data taking at high
luminosity at the LHC and using timing detectors with a good timing resolution
(see Fig.~\ref{signif} right).

In some scenario such as NMSSM where the Higgs boson decays in $h \rightarrow
aa \rightarrow \tau \tau \tau \tau$ where $a$ is the lighter of the two
pseudo-scalar Higgs bosons, the discovery might come only from
exclusive diffractive Higgs production~\cite{higgsnew} ($m_a<2 m_b$ is natural in NMSSM with $m_a>2
m_{\tau}$ somewhat preferred).

\subsection{Photon induced processes at the LHC}

In this section, we discuss  particularly
a new possible test of the Standard Model (SM) predictions
using photon induced processes at the LHC, and especially
$WW$ production~\cite{jochen,olda}. The cross sections of these
processes are computed with high precision using Quantum Electrodynamics
(QED) calculations, and an experimental observation leading to differences with
expectations would be a signal due to beyond standard model effects. The
experimental signature of such processes is the decay products of the $W$ in the
main central detectors from the ATLAS and CMS experiments and the presence of
two intact scattered protons in the final state.

The main source of background is the $W$ pair production
in Double Pomeron Exchange (DPE).
To remove most of the DPE background, it is possible to cut on the $\xi$ of the protons
measured in the proton taggers.
Already with a low integrated luminosity of 200 pb$^{-1}$ it is 
possible to observe 5.6 $W$ pair two-photon events for a background of DPE lower than 0.4,
leading to a signal 
above 8 $\sigma$ for $WW$ production via photon induced processes.

New physics with a characteristic scale (i.e. the typical mass of new particles) 
well above  what can be probed
experimentally at the LHC can manifest itself as a  modification of gauge
boson couplings due to the exchange of new heavy particles. The conventional
way to investigate the sensitivity to the potential new physics is
to introduce an effective
Lagrangian with additional higher
dimensional terms parametrized with anomalous parameters.  
We consider the modification of the $\wwgamma$ triple gauge boson vertex with additional terms conserving $C-$ and $P-$
parity separately, that are parametrized with two anomalous parameters $\dkap$, $\lam$. 
For 30 fb$^{-1}$,
the reach on  $\dkap$ and $\lam$ is respectively 0.043 and 0.034, improving the direct limits
from hadronic colliders by factors of 12 and 4 respectively (with respect to the LEP indirect 
limits, the improvement is only about a factor 2).
Using a luminosity of 200 fb$^{-1}$, present sensitivities
coming from the hadronic colliders
can be improved by about a factor 30, while the LEP sensitivity can be improved
by a factor 5. 

\begin{figure}
\begin{center}   
\includegraphics[width=2.4in]{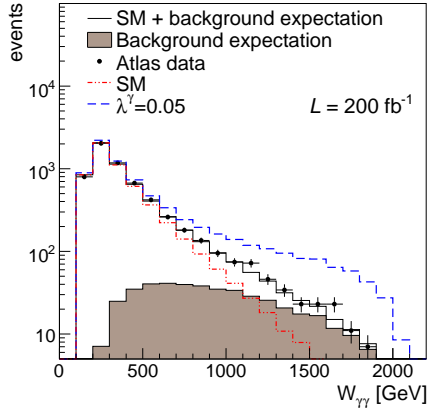}
\caption{Distributions of the $\gamma\gamma$ photon invariant mass $W_{\gamma\gamma}$ 
measured with the forward detectors using $W_{\gamma\gamma}=\sqrt{s\xi_1\xi_2}$. The 
effect of the $\lam$ anomalous parameter appears at high $\gamma\gamma$ 
invariant mass (dashed line).  The SM background is indicated in dot-dashed line, 
the DPE background as a shaded area and their combination
in full line. The black points show the ATLAS data smeared according to a Poisson distribution. }
\label{fig:Wspectrum200}
\end{center}
\end{figure}

It is worth noticing that many observed events are expected in the region $W_{\gamma\gamma}>1$ TeV where 
beyond standard model effects, such as SUSY, new strong dynamics at the TeV
scale, anomalous coupling, etc., are expected (see Fig.~\ref{fig:Wspectrum200}). 
It is expected that the LHC
experiments will collect 400 such events predicted by QED with $W>$1 TeV for a luminosity of 200
fb$^{-1}$ which will allow to probe further the SM expectations. In the same way
that we studied the $WW\gamma$ coupling, it is also possible to study the
$ZZ\gamma$ one. The SM prediction for the $ZZ\gamma$ coupling is 0, and any
observation of this process is directly sensitive to anomalous coupling (the main
SM production of exclusive $ZZ$ event will be due to exclusive Higgs boson
production decaying into two $Z$ bosons if the Higgs boson exists in the
relevant mass range).
The $WW$ cross section measurements are also sensitive to anomalous quartic
couplings~\cite{QuarticCoupling}, and recent studies showed that the sensitivity on quartic
coupling is 10000 times
better than at LEP with only a luminosity of 10 fb$^{-1}$. In addition, it is
possible to produce new physics beyond the Standard Model. 
Two photon production of SUSY leptons as an example has been 
investigated and the cross section for 
$\gamma \gamma \rightarrow \tilde{l}^+ \tilde{l}^-$ can be larger than 1~fb.

\section{The AFP project at the LHC}

\subsection{Motivation}

The motivation to install forward detectors in ATLAS and CMS is quite
clear. Two locations for the forward detectors are considered at
220 and 420 m respectively to ensure a good coverage in $\xi$ or in mass of the
diffractively produced object as we will see in the following. Installing
forward detectors at 420 m is quite challenging since the detectors will be
located in the cold region of the LHC and the cryostat has to be modified to
accomodate the detectors. In addition, the space available is quite small and
some special mechanism called movable beam pipe are used to move the detectors
close to the beam when the beam is stable enough. The situation at 220 m is
easier since it is located in the warm region of the LHC.
The AFP (ATLAS Forward Physics)
project is under discussion in the ATLAS collaboration and includes both 220
and 420 m detectors on both sides of the main ATLAS detector.

The physics motivation of this project corresponds to different domains of
diffraction which we already discussed:
\begin{itemize}
\item A better understanding of the inclusive diffraction mechanism at the LHC 
\item Looking for Higgs boson diffractive production in double pomeron exchange in
the Standard Model or supersymmetric extensions of the Standard Model~\cite{ushiggs,
lavignac} and measuring its properties (mass, spin...). This is
clearly a challenging topic especially at low Higgs boson masses where the
Higgs boson decays in $b \bar{b}$ and the standard non-diffractive search is
difficult. 
\item Sensitivity to the anomalous coupling of the photon by measuring the QED
production cross section of $W$ boson pairs~\cite{olda}.
\end{itemize}

\subsection{Forward detector design and location}
\begin{figure}
\begin{center}
\epsfig{file=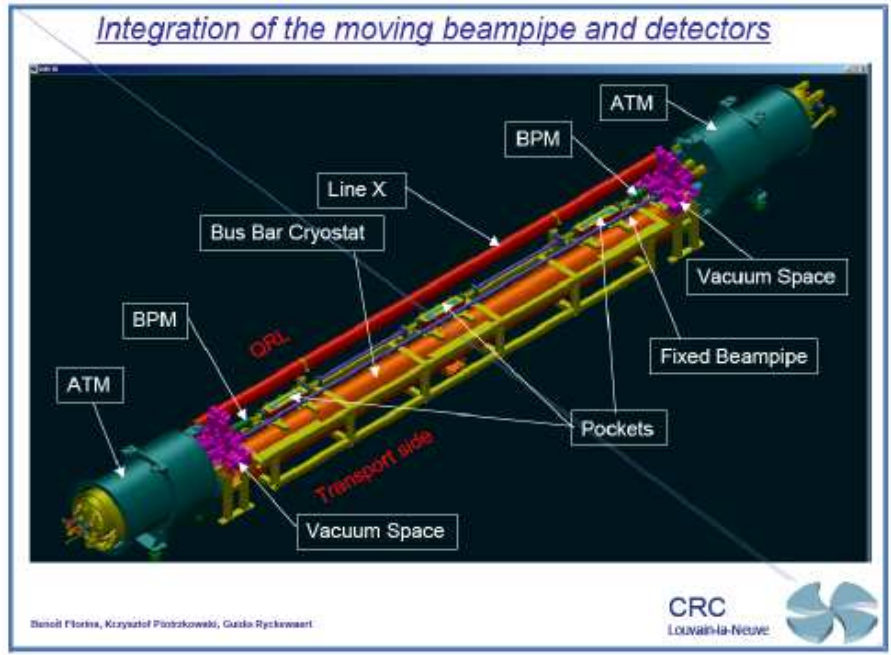,width=12cm}
\caption{Scheme of the movable beam pipe.}
\label{movable}
\end{center}
\end{figure}

As we mentionned in the previous section, it is needed to install movable
beam pipe detectors~\cite{afp} at 220 and 420 m. The scheme of the movable beam pipe
is given in Fig.~\ref{movable}. The principle developed originally for
the ZEUS detector to tag electrons at low angle is quite simple and follows from
the 
same ideas as the roman pots. The beam pipe is larger than the usual one and 
can host the sensitive detectors to tag the diffracted protons in the
final state. When the beam is stable, the beam pipe can move so that the
detectors can be closer to the beam. The movable beam pipe acts in a way as a
single direction roman pot. In Fig.~\ref{movable}, we see the Beam Position
Monitors (BPM) as well as the pockets where the detectors can be put.
The detectors can be aligned and calibrated using the BPMs as well as exclusive
dimuon events. The dimuon mass can be well measured using the central muon
detectors from ATLAS and can be compared to the result obtained using the
missing mass method by tagging the final state proton in the forward detectors.
This allows to calibrate the forward detectors by using data directly.
The exclusive muon production cross section is expected to be high enough to
allow this calibration on a store-by-store basis.

\begin{figure}
\begin{center}
\epsfig{file=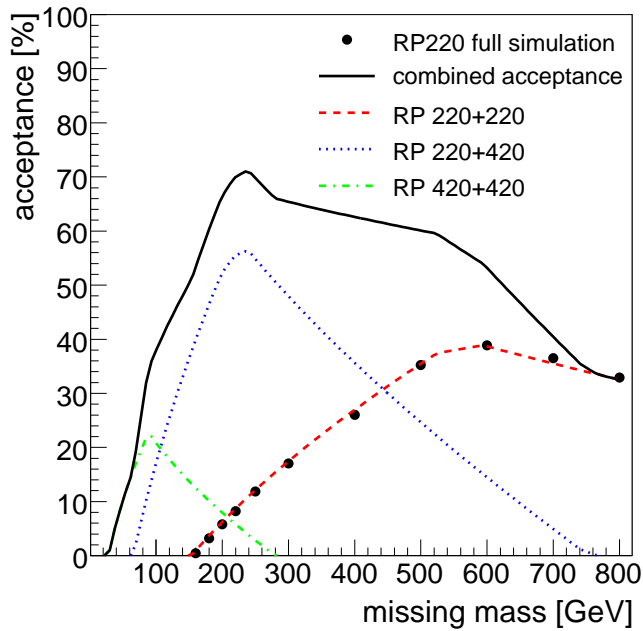,width=9cm}
\caption{Forward detector acceptance as a function of missing mass
assuming a 10$\sigma$ operating positions, a dead edge for the detector of 50
$\mu m$ and a thin window of 200 $\mu m$.}
\label{accept1}
\end{center}
\end{figure}

The missing mass acceptance is given in Fig.~\ref{accept1}. 
The missing mass acceptance using only
the 220 m pots starts at 135 GeV, but increases slowly as a function of missing
mass. It is clear that one needs both detectors at 220 and 420 m to obtain a good
acceptance on a wide range of masses since most events are asymmetric (one tag
at 220 m and another one at 420 m). The precision on mass reconstruction using
either two tags at 220 m or one tag at 220 m and another one at 420 m is of the
order of 2-4 \% on the full mass range, whereas it goes
down to 1\% for symmetric 420 m tags.

\subsection{Detectors inside forward detectors for the AFP project}
We propose to put inside the forward detectors two kinds of detectors, 
namely 3D Silicon
detectors to measure precisely the position of the diffracted protons, and the mass
of the produced object and
$\xi$, and precise timing detectors.

The position detectors will consist in 3D Silicon
detector which allow to obtain a resolution in position better than
10 $\mu$m. The detector is made of 10 layers of 3D Si pixels of 50 $\times$
400 $\mu$m. One layer contains 9 pairs of columns of 160 pixels,
the total size being 7.2 $\times$ 8 mm$^2$. 
The detectors will
be read out by the standard ATLAS pixel chip~\cite{afp}. The latency time of the chip is 
larger than 6 $\mu$s which gives enough time to send back the local L1 decision from
the forward detectors to ATLAS (see the next paragraph about trigger for more detail),
and to receive the L1 decision from ATLAS, which means a distance of about 440
m. It is also foreseen to perform a slight modification of the chip to
include the trigger possibilities into the chip.

The timing detectors are necessary at the highest luminosity of the LHC to
identify from which vertex the protons are coming from. It is expected that up
to 35 interactions occur at the same bunch crossing and we need to identify from
which interaction, or from which vertex the protons are coming from. A precision
of the order of 1 mm or 2-5 ps is required to distinguish between the
different vertices and to make sure that the diffracted protons come from the
hard interactions. Picosecond timing detectors are still a challenge and are
developped for medical and particle physics applications. 
Two technogies are developped, either using as a radiator --- with the aim
to emit photons by the diffracted protons ---- or gas (gas Cerenkov detector or GASTOF) 
or a crystal of about 2.5 cm (QUARTIC), and the signal can be read out by 
Micro-Channel Plates Photomultipliers~\cite{afp}. 
The space resolution of those detectors should be of the
order of a few mm since at most two protons will be detected in those detectors
for one given bunch crossing at the highest luminosity. The detectors can be read
out with a Constant Fraction Discriminator which allows to improve the timing
resolution significantly compared to usual electronics.
A first version of the
timing detectors is expected to be ready in 2010 with a resolution of
20-30 ps, and the final version by 2012-2013 with a resolution of 2-5 ps.
The phototube aging due to the high proton rate at LHC at high luminosity is still
an opened issue which is being solved in collaboration with the industry.

\subsection{Trigger principle and rate}
In this section, we would like to give the principle of the trigger using the
forward detectors at 220 m as well as the rates obtained using a simulation of the
ATLAS detector and trigger framework~\cite{afp}.

\begin{figure}
\begin{center}
\epsfig{file=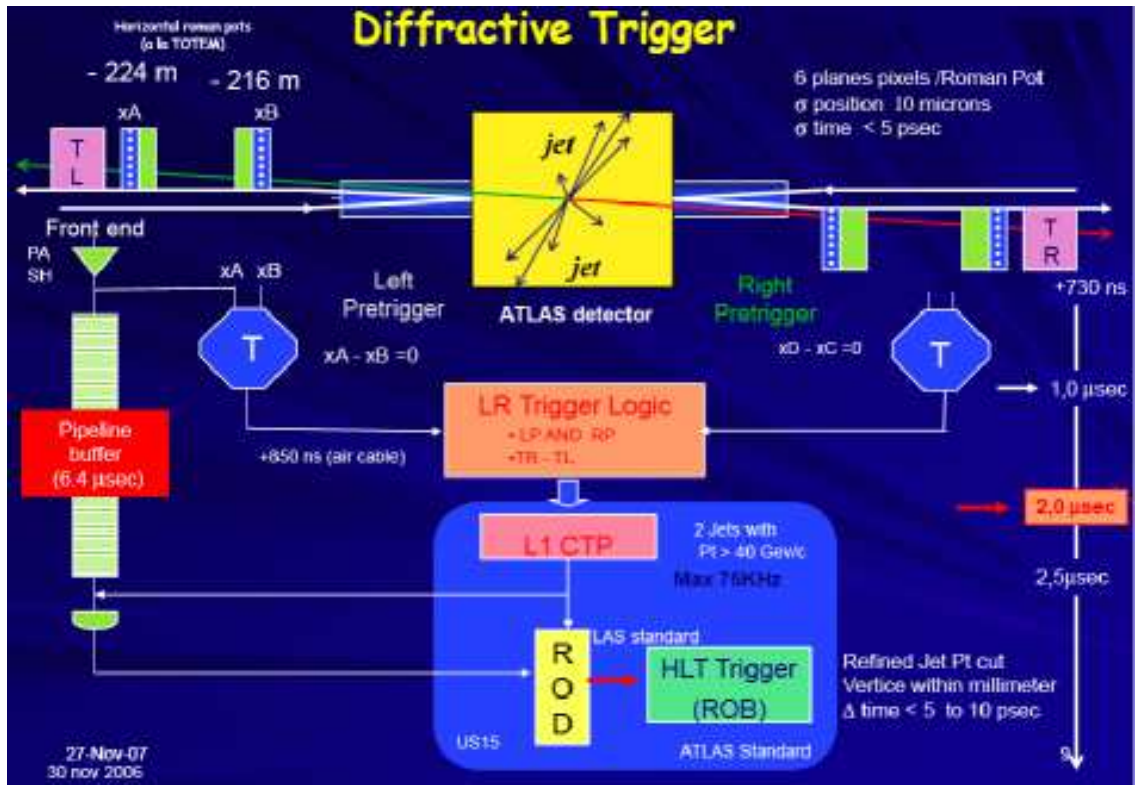,width=15cm}
\caption{Scheme for L1 trigger for the AFP project.}
\label{trig}
\end{center}
\end{figure}

The principle of the trigger is shown in Fig.~\ref{trig} in the case of a Higgs boson
decaying into $b \bar{b}$ as an example.
The first level trigger comes directly from two different 3D Silicon layers
in each forward detector. It is more practical to use two dedicated planes for
triggering only since it allows to use different signal thresholds for trigger
and readout. The idea is to send at most five strip addresses which are hit
at level 1 (to simplify the trigger procedure, we group all pixels in 
vertical lines as one element only for the trigger
since it is enough to know the distance in the horizontal direction to have 
a good approximation of $\xi$). 
A local trigger is defined at the movable beam pipe level on each side of
the ATLAS experiment by combining the two trigger planes in each movable
beam pipe and
the forward detectors as well. If the hits are found to be compatible (not issued by
noise but by real protons), the strip addresses are sent to ATLAS, which allows
to compute the $\xi$ of each proton, and the diffractive mass. This information
is then combined with the information coming from the central ATLAS detector,
requesting for instance two jets above 40 GeV in the case shown in Fig.~\ref{trig}. At
L2, the information coming from the timing detectors for each diffracted proton
can be used and combined with the position of the main vertex of ATLAS to check
for compatibility. Once a positive ATLAS trigger decision is taken (even without
any diffracted proton), the readout informations coming from the 
forward detectors are sent to ATLAS as any subdetector.

The different trigger possibilities for the forward detectors are given below:
\begin{itemize}
\item {\it {\bf Trigger on DPE events at 220 m:}}
This is the easiest situation since two protons can be requested at Level 1 
at 220 m. 
\item {\it {\bf Trigger on DPE events at 220 and 420 m:}}
This is the most delicate scenario since the information from the 420 m pots cannot
be included at L1 because of the L1 latency time of ATLAS. 
The strategy (see Table 1) is to
trigger on heavy objects (Higgs...) decaying in $b \bar{b}$ by requesting
a positive tag (one side only) at 220 m with $\xi < 0.05$
(due to the $420\mathrm{m}$ RP acceptance
in  $\xi$, the proton momentum fractional loss in the $220\mathrm{m}$ forward
detector cannot be too high
if the Higgs mass is smaller than $140\,\mathrm{GeV}$), and topological cuts
on jets such as the exclusiveness of the process
($(E_{jet1}+E_{jet2})/E_{calo}>0.9$, 
$(\eta_1+\eta_2)\cdot\eta_{220} > 0$, where 
$\eta_{1,2}$ are the pseudorapidities of the two L1 jets, and $\eta_{220}$
the pseudorapidity of the proton in the $220\mathrm{m}$ movable beam pipe). 
This trigger can hold without prescales to a
luminosity up to 2.10$^{33}$ cm$^{-2}$s$^{-1}$, but would require an upgrade of the
ATLAS L1 trigger. In addition, we are still looking at new possibilities
to trigger in the same channels at higher luminosities. The ideas might
be to use the layer 0 silicon or to the fact that $b$ jets are thinner than gluon
jets allowing a smaller sliding window.
Let us note that the rate will be of the order of a few Hz at L2 by adding a cut on
a presence of a tag in the 420 pots, on timing, and also on the compatibility of
the rapidity of the central object computed using the jets or the protons in
the forward detectors.
\end{itemize}
The trigger on $W$, top...will be given by ATLAS directly without any special forward
trigger.

\begin{footnotesize}
\begin{table}
\begin{center}
\begin{tabular}{|c|c|c|c|c|c|}
\hline
${\cal L}$ & $n_{pp}$ per & 2-jet  & RP200 
& $\xi < 0.05$  & Jet \cr
$E_T > 40\,\mathrm{GeV}$ & bunch & rate $[\mathrm{kHz}]$ &
reduction & reduction & Prop.  \cr
 & crossing & $[\mathrm{cm}^{-2}\cdot\mathrm{s}^{-1}]$& factor & factor & \cr
\hline
$1\times10^{32}$ & 0.35 & 2.6 & 120 & 300 & 1200 \cr
$1\times10^{33}$ & 3.5 & 26 & 8.9 & 22 & 88 \cr
$2\times10^{33}$ & 7 & 52 & 4.2 & 9.8 & 39.2 \cr
$5\times10^{33}$ & 17.5 & 130 & 1.9 & 3.9 & 15.6 \cr
$1\times10^{34}$ & 35 & 260 & 1.3 & 2.2 & 8.8 \cr
\hline
\end{tabular}
\end{center}
\caption{L1 rates for 2-jet trigger with $E_T > 40\,\mathrm{GeV}$ and
additional reduction factors due to the requirement of triggering on
diffractive proton at $220\,\mathrm{m}$, and also on jet properties.
The total rate should not exceed a few kHz at L1.}
\label{t_trigger}
\end{table}
\end{footnotesize}

\section{Conclusion}
In this review, we described briefly the results on diffraction from Tevatron 
stressing in particular the search for diffractive exclusive events. This kind of events
together with the QCD studies and the search for anomalous $\gamma W$ couplings motivated
the project to install forward detectors in the ATLAS and CMS collaborations.
These studies are only a small part of the QCD studies which can be performed at the
LHC concerning PDF~\cite{ourreview} or low $x$ resummation effects especially
in the Mueller-Navelet or jet gap jet channels~\cite{bfkl}.



\begin{thebibliography}{99}

\bibitem{ourreview} M. Boonekamp, F. Chevallier, C. Royon, L. Schoeffel,
ArXiv:0902.1678;   C.~Royon,
  Acta Phys.\ Polon.\  B {\bf   37} (2006) 3571; 
  Acta  Phys. Polon. B {\bf 39} (2008) 2339.; H. Abramowicz, contribution given
at this conference.

\bibitem{heraf2d} H1 Collaboration, Eur. Phys. J. {\bf C48}
(2006) 715; Eur. Phys. J. {\bf C48}
(2006) 749; ZEUS Collaboration,
Nucl. Phys. {\bf  B 713} (2005) 3.


\bibitem{dglap} 
G.Altarelli and G.Parisi,
{\it Nucl. Phys.} {\bf B126}  18C (1977) 298;
V.N.Gribov and L.N.Lipatov, {\it Sov. Journ. Nucl. Phys.} (1972) 438 and 675;
Yu.L.Dokshitzer, {\it Sov. Phys. JETP.} {\bf 46} (1977) 641.

\bibitem{f2dfit}
C. Royon, L. Schoeffel, J. Bartels, H. Jung, R. Peschanski, Phys. Rev. {\bf D63}
(2001) 074004;
 C.~Royon, L.~Schoeffel, S.~Sapeta, R.~B.~Peschanski and E.~Sauvan,
  Nucl.\ Phys.\  B {\bf 781} (2007) 1 ; 
 C.~Royon, L.~Schoeffel, R.~B.~Peschanski and E.~Sauvan,
  Nucl.\ Phys.\  B {\bf   746} (2006) 15.


\bibitem{chic} M. Rangel, C. Royon, G. Alves, J. Barreto, R. Peschanski, Nucl. Phys. B{\bf 774}
(2007) 53.

\bibitem{cdfrjj} CDF Collaboration, Phys. Rev. {\bf D77} (2008) 052004.


\bibitem{oldab} O.Kepka, C. Royon, Phys. Rev.D{\bf 76} (2007) 034012.

\bibitem{ushiggs} C.~Royon,
  Mod.\ Phys.\ Lett.\ A {\bf 18}, 2169 (2003) and references therein;
M. Boonekamp, R. Peschanski, C. Royon, Phys. Rev. Lett. {\bf  87 } 
(2001) 
251806; Nucl. Phys. {\bf B669} (2003) 277;
M. Boonekamp, A. De Roeck, R. Peschanski, C. Royon, Phys. Lett.  {\bf  
B550} (2002) 93;
V.A. Khoze, A.D. Martin, M.G. Ryskin, Eur. Phys. J. {\bf C19} (2001) 477;
Eur. Phys. J. {\bf C23} (2002) 311;
Eur. Phys. J. {\bf C24} (2002) 581; arXiv:0802.0177;
Phys. Lett. B650 (2007) 41; A.B. Kaidalov, V.A. Khoze, A.D. Martin, M.G. Ryskin,
Eur. Phys. J. {\bf C33} (2004) 261; Eur. Phys. J. {\bf C31} (2003) 387;
R. Peschanski, M. Rangel, C. Royon, preprint arXiv:0808.1691. 

\bibitem{cdfgamma} CDF Collaboration, Phys. Rev. Lett. 99 (2007) 242002. 

\bibitem{afp} FP420 Coll., see http://www.fp420.com;
ATLAS and CMS TDR to be submitted; see: 
http://project-rp220. web.cern.ch/project-rp220/index.html; C. Royon,
preprint arXiv:0706.1796, proceedings of 15th International 
Workshop on Deep-Inelastic Scattering and Related Subjects (DIS2007), 
Munich, Germany, 16-20 Apr 2007. 

\bibitem{lavignac} 
M. Boonekamp, J. Cammin, S. Lavignac, R. Peschanski, C. Royon,
Phys. Rev. {\bf D73} (2006) 115011, and references therein.

\bibitem{higgsnew} B. Cox, F. Loebinger, A. Pilkington, JHEP 0710 (2007) 090;
S. Heinemeyer et al., Eur.~Phys.~ J. C {\bf 53} (2008) 231;
J.R. Forshaw, J.F. Gunion, L. Hodgkinson, A. Papaefstathiou, 
A.D. Pilkington, JHEP {\bf 0804} (2008) 090.

\bibitem{jochen} 
M. Boonekamp, J. Cammin, R. Peschanski, C. Royon, Phys. Lett.
{\bf B654} (2007) 104.

\bibitem{olda} 
O. Kepka, C. Royon, Phys. Rev. D {\bf 78} (2008) 073005. 

\bibitem{QuarticCoupling} E. Chapon, F. Chevallier, O. Kepka, C. Royon, in preparation;
  T.~Pierzchala and K.~Piotrzkowski,
  arXiv:0807.1121 [hep-ph]; N.~Schul and K.~Piotrzkowski,
  arXiv:0806.1097 [hep-ph].



\bibitem{bfkl}H.Navelet, R.Peschanski, Ch.Royon, S.Wallon,
Phys. Lett. {\bf B385} (1996) 357; H.~Navelet, R.~Peschanski, C.~Royon,
Phys. Lett. B366 (1996) 329;
A.~Bialas, R.~Peschanski, C. Royon,
Phys. Rev. D {\bf 57} (1998) 6899; S.~Munier, R.~Peschanski, C.~Royon, 
Nucl. Phys. B {\bf 534} (1998) 297;
O. Kepka, C. Marquet, R. Peschanski and C. Royon, 
Phys. Lett. {\bf B655} (2007) 236; Eur. Phys. J.
{\bf C55} (2008) 259; C. Marquet, C. Royon, Nucl. Phys. {\bf B739} (2006) 131;
Phys. Rev. {\bf D79} (209) 034028;
G.~Beuf, R.~Peschanski, C.~Royon, D.~Salek, Phys. Rev. D78 (2008) 074004;
F. Chevallier, O. Kepka, C. Marquet, C. Royon,
arXiv:0903.4598. 


\end{thebibliography}
\end{document}